# Martensite structures and twinning in substrate-constrained epitaxial Ni-Mn-Ga films deposited by a magnetron co-sputtering process

Jérémy Tillier[a, b, *], Daniel Bourgault[a], Sébastien Pairis[a], Luc Ortega[a], Nathalie Caillault[b], Laurent Carbone[b]

[a]*Institut NÉEL/CRETA, CNRS et Université Joseph Fourier, BP166, F-38042 Grenoble Cedex 9, France*
[b]*Schneider Electric France, 38TEC/T1, F-38050 Grenoble Cedex 9, France*

**Abstract**

In order to obtain Ni-Mn-Ga epitaxial films crystallized in martensite structures showing Magnetic-Induced Rearrangement (MIR) of martensite variants, a fine control of the composition is required. Here we present how the co-sputtering process might be helpful in the development of Ni-Mn-Ga epitaxial films. A batch of epitaxial Ni-Mn-Ga films deposited by co-sputtering of a Ni-Mn-Ga ternary target and a pure manganese target has been studied. The co-sputtering process allows a precise control of the film compositions and enables keeping the epitaxial growth of Ni-Mn-Ga austenite during deposition at high temperature. It gives rise to tune the content of the MIR-active 14-modulated martensite in the film at room temperature, as well as micro and macro-twinned domains sizes.

*PACS:* 68.55.Nq; 81.15.Kk; 81.15.Cd

*Keywords:* Ni-Mn-Ga Magnetic Shape Memory Alloy, Thin film, Epitaxial growth, Deposition by co-sputtering, Structure

## 1. Introduction

The Magnetic Shape Memory Alloys (MSMA) constitute a new class of Shape Memory Alloys (SMA). In addition to the Martensitic Transformation (MT), these metallic alloys show interesting magnetic properties. In Ni-Mn-Ga, the magneto-structural coupling between magnetic moments and martensite variants leads to a large panel of properties like Magnetic Induced Martensite (MIM) [1] or Magnetic Induced Rearrangement (MIR) of martensite variants [2]. This last effect has gained considerable attention since magnetic induced strains up to 10% have been observed in Ni-Mn-Ga single crystals [3-5].

Today, large efforts are carried out to develop MIR-active Ni-Mn-Ga films because of their promising applications as new micro-actuators or sensors for Micro-Electro-Mechanical Systems (MEMS) [2, 6-13]. Epitaxial growth is considered to be the most hopeful process because the highest strains have been reported in bulk single crystals. Moreover, evidences of MIR in films have been only reported for epitaxially deposited layers to date [2, 11-13]. A key requirement to obtain MIR is that the magnetically induced stress exceeds the mechanical stress needed to rearrange the structure. Large magneto-crystalline anisotropy and low twinning-stress of the martensite phase are thus prerequisites. These conditions are only fulfilled in modulated-structures like the ten-modulated (10M) or the fourteen-modulated (14M) martensites. The twinning stress of the non-modulated (NM) martensite is too high to allow MIR [14]. At Room Temperature (RT), the martensite structure has been found to be strongly dependent on the alloy composition, the proportion of NM-martensite increasing with the average valence electron concentration per atom (e/a) [14].

Sputtering process has been proved to be an efficient technique to deposit reproducible epitaxial films on various substrates. Tailoring the composition has been realized by using different techniques like changing the target composition [10], the deposition temperature [11], the sputtering reactor pressure [15] or applying a negative bias voltage on the substrate [16]. Varying the target composition is a quite simple route but preferential sputtering of Ni atoms occurs during the process [17]. It is thus difficult to finely adjust the film composition using this preparation route.

Here we present a process allowing a precise tuning of the composition of Ni-Mn-Ga films without changing the above-mentioned sputtering parameters. We use the simultaneous deposition from two targets: one made of the ternary Ni-Mn-Ga alloy, the other being of pure manganese. All parameters have been fixed to facilitate epitaxial growth on (001) MgO monocrystalline substrates and the composition has been adjusted by changing the applied power on the manganese target. Influence of the composition on the structure of the martensite phases and on the surface-morphology is discussed in connection with the twinning phenomena. This demonstrates the powerfulness of the magnetron co-sputtering process for the development of Ni-Mn-Ga epitaxial films.

## 2. Experimental

Epitaxial Ni-Mn-Ga films have been deposited by the magnetron sputtering technique. A low residual pressure in the range of $10^{-6}$ Pa was used to avoid any oxidation of the films. Epitaxial films have been grown in a confocal plasma reactor equipped with six cathodes: three operated in Direct Current (DC), three operated at Radio Frequency (RF). In the following, we use simply two of these. The ternary $Ni_{56}Mn_{22}Ga_{22}$ alloy was inserted on a DC-cathode whereas the pure Mn target was inserted on a RF-cathode to enable tuning of the composition. Each target had a 2" diameter and was located 0.105 m above the substrate. In order to ensure chemical homogeneity and constant film thicknesses, the depositions were made with the substrate rotating at a speed of 5 rpm. The (001) MgO monocrystalline substrate temperature has been fixed at 773 K and the applied power (86 W) on the ternary target was optimized to obtain a deposition rate of 1 µm/h. This (substrate temperature)-(deposition rate) couple of parameters give rise to a Ni-Mn-Ga(001)[110] // MgO(001)[100] epitaxial relationship between the deposited austenite and the substrate (see [18] for more details). The argon gas with a high purity of 99.999 vol% was filled into the chamber with a flow of 35 Standard Cubic Centimeters per minute (SCCM). The sputtering pressure was kept at 0.8 Pa by controlling the angular position of the valve insulating the turbo-molecular pump from the chamber.

The film and target compositions were determined by Energy Dispersive X-ray spectroscopy (EDX) using a JEOL 840A Scanning Electron Microscope (SEM). The EDX measurements have been performed without Ni-Mn-Ga standard, leading to an accuracy on the quantitative compositions of 2 at.%. The accuracy on the relative compositions is less than 0.5%. This has been verified by realizing three EDX analyses in spot mode and three EDX analyses in average mode for each film. The analyses in average mode were performed using a magnification of 1 kX. Structural characterizations were made using X-Ray Diffraction (XRD). Measurements have been carried in a four-circle instrument (Seifert MZ IV) using the copper $K_\alpha$ radiation and with an optic (Xenocs) at the output of the source allowing a low horizontal divergence (0.06°) of the X-ray beam. The diffractometer was equipped with a rear mono-chromator in order to enhance the signal to noise ratio. Alignments of the samples have been realized using the (002) reflection of the MgO substrate. It allows probing crystallographic orientations of all Ni-Mn-Ga phases in absolutes coordinates, the single crystalline substrate acting as a reference system [19]. θ-2θ scans have been measured for tilt angle ψ ranging from zero to ten degrees at two in-plane rotation angles φ selected with respect to the known epitaxial relationship between the substrate and austenite. The microstructures were analyzed using various microscopy techniques. Optical microscopy with polarized light (Zeiss microscope equipped with a Hamamatsu ORCA-ER digital camera) has been used to investigate the surface macro-twinned domains. The surface micro-twins have been characterized using the secondary electron (SE) detector of a Field-Emission Scanning Electron Microscope (FESEM Zeiss Ultra +: 3kV, 3mm, 10kX).

## 3. Compositions versus RF-power on manganese target

The multi-target sputtering reactor allows an independent control of the power applied on each target. In order to optimize the DC-power applied on the ternary target, different powers have been tested for a fixed deposition time of one hour and film thicknesses have been measured by profilometry (not shown). For the deposition parameters mentioned in the experimental, the ternary $Ni_{56}Mn_{22}Ga_{22}$ target gives rise to $Ni_{60}Mn_{20}Ga_{20}$ films. The difference of composition between the target and the film can be ascribed to the fact that preferential evaporation of Mn and Ga occurs during deposition at high temperature [20]. The Ni-rich $Ni_{60}Mn_{20}Ga_{20}$ composition leads to a high average valence electron concentration per atom (e/a) slightly above 8.0. In bulk alloys showing well defined inter-martensitic transitions, this (e/a) factor is thought to be responsible of NM martensite at room temperature [14].

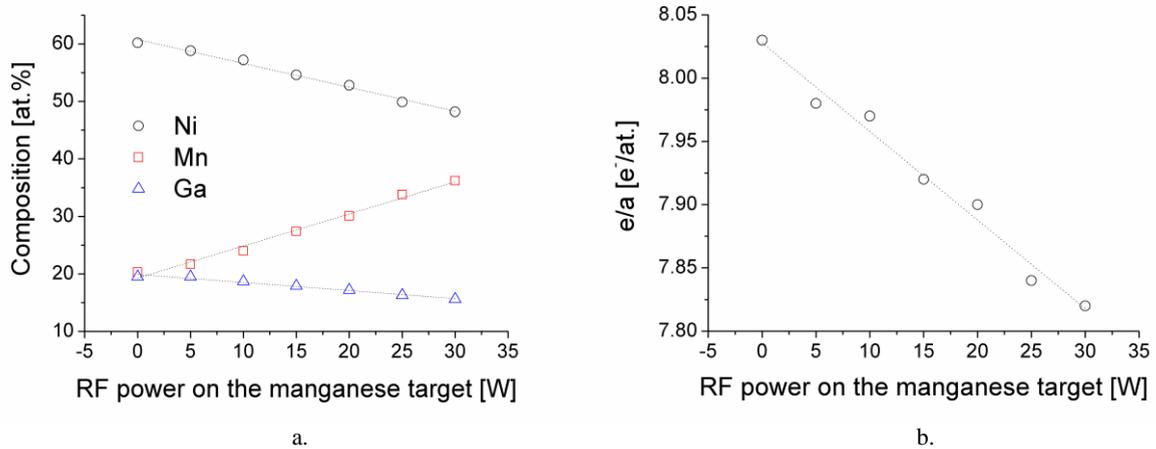

a.   b.

Figure 1.a.: Elemental compositions as a function of the RF-power applied on the manganese target. Atomic contents of nickel, manganese and gallium are represented with open-circles, open-squares and open-triangles, respectively. Figure 1.b.: Influence of the manganese target RF-power on the average valence electron concentration per atom (e/a).

Figure 1.a. represents the elemental compositions Ni, Mn and Ga versus the RF-power applied on the manganese target. The flux of sputtered manganese atoms increases with the power applied on the manganese target. It leads to rising up the deposited manganese atomic content whereas the proportions of nickel and gallium decline. Moreover, figure 1 reveals that the decrease of nickel content is more pronounced than that of gallium. In fact, the composition of films deposited without power on the manganese target is $Ni_{60}Mn_{20}Ga_{20}$. This composition contains three times more nickel than gallium. The proportion of nickel is thus expected to decline three times faster than the gallium content when applying a power on the manganese target. The linear regressions demonstrate the decline is 0.42at.%/W for Ni and 0.14at.%/W for Ga, agreeing with the previous argument. This particular feature leads to a linear decrease of the average electron concentration per atom when increasing the applied power on the manganese target, as shown in figure 1.b. It varies from around 8.0 (no power on the manganese target) down to around 7.8 (30W on the manganese target) with a negative slope of 0.007 (e$^-$/at.)/W. An increase of 14M-martensite content at RT might consequently be expected for increasing manganese target powers.

## 4. Compositions versus martensite structures

Thomas et al. [11] have demonstrated that in the case of Ni-Mn-Ga films epitaxially deposited on MgO substrates, a thin austenitic layer persists on the film/substrate interface, even at temperatures below the martensitic finish temperature of the film volume. The presence of this interfacial austenite layer is due to substrate-induced constrains which hinder the martensitic transformation at the substrate interface. In fact, the epitaxial relationship of the deposited austenite (A) on (001)MgO is A(001)[110]//MgO(001)[100], giving rise to a crystallographic mismatch $(\sqrt{2}a_{MgO} - a_A)/a_A$ around 2%. The martensitic transformation from the parent austenite epitaxially deposited during deposition at high temperature to the martensite phase must accommodate the strain induced by the crystal lattice mismatch between the two phases. The transformation path requires an invariant habit plane along which each cell rotates to restore crystal-lattice continuity across the boundary plane [21]. In order to identify the martensite phases and determine their orientations compared to that of the substrate, θ-2θ XRD-scans were measured in the 2θ-range of (400) reflections of Ni-Mn-Ga martensites for tilt angles ψ ranging from 0° to 10°. The in-plane rotation angles φ have been selected with respect to the known A/MgO epitaxial relationship and twinning planes of Ni-Mn-Ga martensites, which are of (101) type for the bct cell.

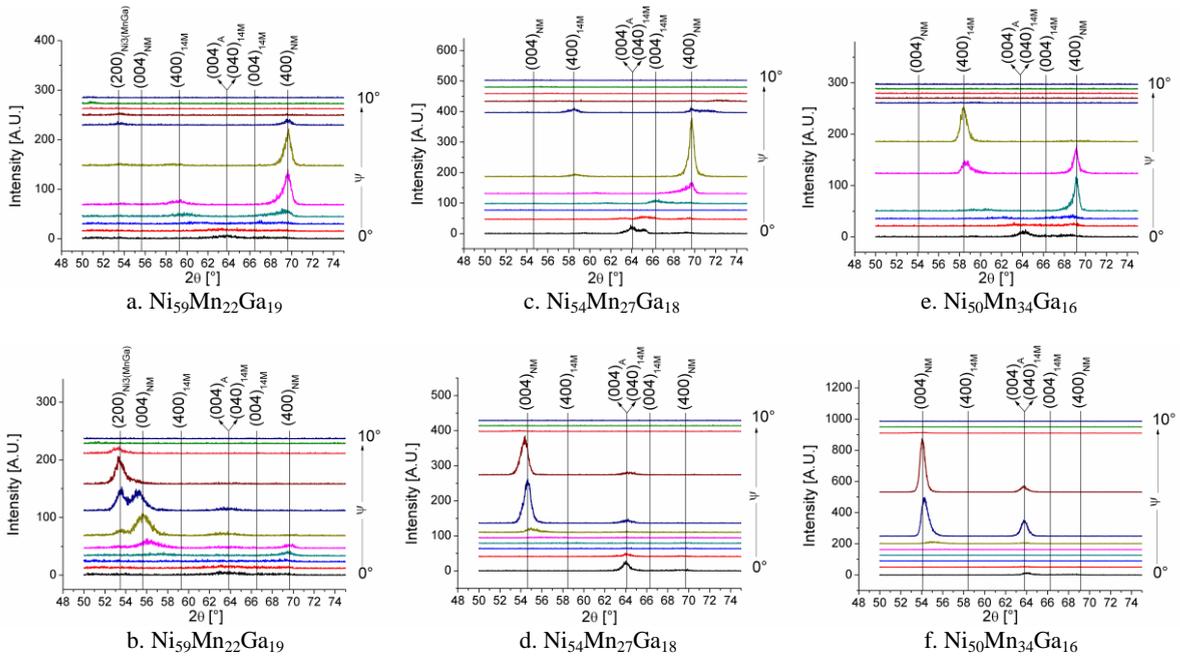

Figure 2: Ψ-dependent XRD θ-2θ scans of three selected samples. The compositions are indicated on the figure. XRD-scans of figures 2.a, 2.c. and 2.e. were realized with the direction of X-Ray beam at 45° to the (100) and (010) edges of the MgO substrates. XRD-scans of figures 2.b., 2.d. and 2.f. were measured with the direction of X-ray beam parallel to the (100) and perpendicular to (010) edges of the MgO substrates.

Figure 2 represents the ψ dependent θ-2θ XRD-scans at selected rotation angles of three selected films deposited for manganese target powers of 5W, 15W and 25W. These powers lead to $Ni_{59}Mn_{22}Ga_{19}$, $Ni_{54}Mn_{27}Ga_{18}$ and $Ni_{50}Mn_{34}Ga_{16}$ compositions, respectively. For all compositions, figure 2 demonstrates the presence of (400) martensite reflections at tilt positions, which are expected from other works on epitaxial films on MgO [10, 11 and 19]. Thus the co-sputtering process does not affect epitaxial growth of austenite at high temperature. All XRD-scans series show the coexistence of three phases: austenite, 14M-martensite and NM-martensite coexisting at room temperature, even for the $Ni_{60}Mn_{20}Ga_{20}$ composition (not shown), where the (e/a) factor is maximal, i.e. around 8.0. The presence of 14M-martensite even at compositions that are expected to give only NM bulk alloys is of particular interest. In fact, due to crystallography, the NM cells cannot directly accommodate on the underlying austenite by simple twinning. As discussed in reference [11], accommodation of the martensite on the austenite thin layer occurs by twinning of the 14M martensite, which possesses a crystallographic axes b equal to that of the austenite cell parameter. Only four of the six possible variants can adapt on the underlying austenite. The two 14M variants with b-axis pointing out-of-plane are not allowed by crystallography. According to the previous argument, no (040) reflections of 14M should be observed in ψ dependent θ-2θ XRD-scans. 14M (040) peaks which are tilted from the substrate normal by 6° to 7° are indeed observed for all the samples. This discrepancy can be explained by the formation of secondary twins along the film thickness.

Figure 2.b. reveals the presence of an additional peak at 2θ = 53.45° for the $Ni_{59}Mn_{22}Ga_{19}$ sample. This peak can be indexed with a (200) reflection of a cubic $Ni_3Mn_xGa_{1-x}$ secondary phase with a cell parameter of around 3.42 Å. Such Ni-rich secondary phases have already been observed in epitaxial Ni-Mn-Ga films [20]. The orientation of the (200) planes of the secondary phase, which is quite similar to that of the (004) planes of the NM-martensite is not fully understood yet. The formation of $Ni_3Mn_xGa_{1-x}$ secondary phases is only observed for the samples deposited for 0, 5 and 10W on the manganese target i.e. such secondary phases only form for film compositions exhibiting a large excess of nickel.

Figure 2 highlights that the 14M-content in the film strongly depends on the film composition, which was precisely tuned by the co-sputtering process. The co-sputtering process allows increasing the content of the MIR-active 14M martensite, as verified by using integrated peak intensities of each phase (not shown).

## 5. Compositions versus twin morphologies

Figure 3 presents the twin surface-morphologies, at different scales, reveled by polarized light and by FESEM images taken in SE mode.

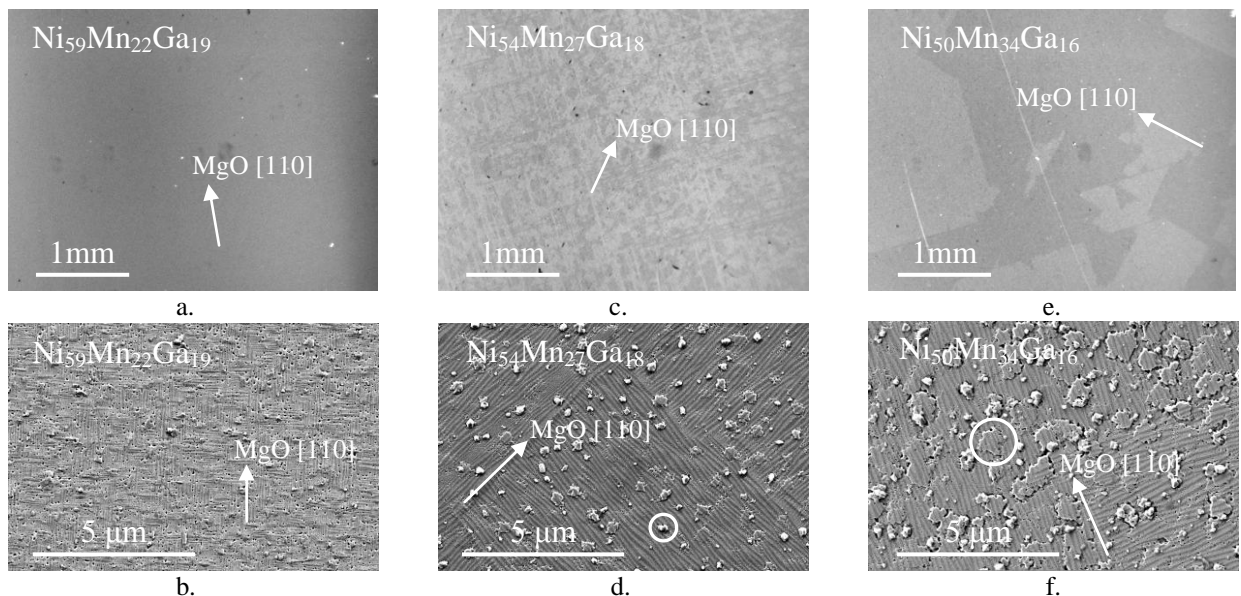

Figure 3: Surface-morphologies of macro-domains revealed by polarized light (figures 3.a., 3.c. and 3.e.) and surface-morphologies of micro-twins by FESEM in SE mode (figures 3.b., 3.d. and 3.f.). Film compositions and scales are indicated on the figure.

Figure 3.a. shows that no macro-twinned domains are resolved by polarized-light for the $Ni_{59}Mn_{22}Ga_{19}$ film. It is due to the low sizes of macro-twinned domains for this composition. In fact, as highlighted by figure 3.b., the macro-twins of the $Ni_{59}Mn_{22}Ga_{19}$ sample are composed of a small number of micro-twins. It explained why no contrast is observed on figure 3.a. Figures 3.c. and 3.e. bring to light the presence of macro-twinned domains at the surface level. It should be noted that branching between two macro-domains occurs in MgO [110] type directions. Figures 3.a., 3.c. and 3.e. clearly bring to light that the macro-twinned domain sizes are affected by the composition, which was precisely tuned with our co-sputtering process.

Figure 3.b., 3.d. and 3.f. display the micro-twinned domains characterizations, which were carried out by FESEM in SE mode in order to reveal the surface topography. Figure 3.b., 3.d. and 3.f. demonstrate that all the film surfaces exhibit pores and terraces (see figure 3.f., a terrace is surrounded for more clarity). For all the films, the FESEM images of figures 3.b., 3.d. and 3.e. also show areas with a clearer contrast (see the area surrounded in white in figure 3.d.). At first glance, such areas might be identified as precipitates or impurities. The number and sizes of such areas increase with the power applied on the manganese target, suggesting that it might be linked to the purity of the manganese target. Nevertheless, no composition variations have been observed by EDX spot-analyses as well as EDX-mapping of the film (not shown). Moreover, the content of impurities in the manganese target is only 0.05 at.%, being mainly of sulphur (240 ppm). It is more likely that such areas with clearer contrast belong to out-grown grains. In fact, the deposition speed increases from 1 µm/h (0W on the manganese target) to 1.2 µm/h for (30W on the manganese target), explaining why the content of out-grown grains increases with the power applied on the manganese target.

## 6. Conclusions

A batch of epitaxial Ni-Mn-Ga films deposited on single-crystalline MgO by co-sputtering of a Ni-Mn-Ga ternary target and a pure manganese target has been studied. The co-sputtering process allows a precise control of the film compositions, keeping the epitaxial growth of Ni-Mn-Ga austenite during deposition at high temperature. By varying the applied RF-power on the manganese target, film compositions were varied from $Ni_{60}Mn_{20}Ga_{20}$ (no power on the manganese target) to $Ni_{48}Mn_{36}Ga_{16}$ (30W on the manganese target). XRD four-circle measurements have revealed that all the films are crystallized in the martensitic state at room temperature. Every sample of the studied batch exhibits co-existence of an austenite layer, 14M-martensite and NM-martensite. Results demonstrate that the 14M-content of the film, which is a MIR-active martensite, increases with the applied power on the manganese target. Microstructure characterizations highlight that both micro-twins and macro-twinned domains morphologies are affected by the composition change, the sizes of both type of twin increasing with the power applied on the manganese target. Thus, presented results demonstrate how helpful might be the co-sputtering process in Ni-Mn-Ga epitaxial films development.

## Acknowledgments

The authors gratefully acknowledge Dr. Philippe Odier for helpful discussions.